\documentclass[sigconf,nonacm]{acmart}

\setcopyright{none}
\renewcommand\footnotetextcopyrightpermission[1]{}

\title{Research on Domain Information Mining and Theme Evolution of Scientific Papers}

\author{Changwei Zheng}
\author{Zhe Xue}
\authornote{Corresponding author.}
\author{Meiyu Liang}
\author{Feifei Kou}
\author{Zeli Guan}
\affiliation{%
  \institution{School of Computer Science (National Pilot School of Software Engineering), Beijing University of Posts and Telecommunications}
  \city{Beijing}
  \country{China}
}

\begin{document}

\begin{abstract}
In recent years, with the increase of social investment in scientific research, the number of research results in various fields has increased significantly. Cross-disciplinary research results have gradually become an emerging frontier research direction. There is a certain dependence between a large number of research results. It is difficult to effectively analyze today's scientific research results when looking at a single research field in isolation. How to effectively use the huge number of scientific papers to help researchers becomes a challenge. This paper introduces the research status at home and abroad in terms of domain information mining and topic evolution law of scientific and technological papers from three aspects: the semantic feature representation learning of scientific and technological papers, the field information mining of scientific and technological papers, and the mining and prediction of research topic evolution rules of scientific and technological papers.
\end{abstract}




\maketitle

\noindent\textbf{CLC number}\quad TP391

In recent years, with the rapid development of science and technology, researchers have gradually deepened their research in various fields, and the number of scientific and technological papers has exploded. Rich scientific and technological paper resources can effectively help all kinds of scientific researchers, enterprises and other scientific and technological innovation. However, in the face of a huge number of paper resources, how to effectively manage and quickly obtain the resources needed by researchers has become an urgent and critical issue that needs to be solved urgently\cite{hu2019knowledgegraphplatform}.

The division of subject areas is of great significance for the research and analysis of scientific and technological paper data, and the dissertation database is an important source for scholars to obtain the latest scientific research results. However, due to the wide variety of thesis databases, the classification methods and standards adopted by different institutions are different, and the staff often needs to rely on a lot of professional knowledge to manually classify papers when including papers. Such manual classification undoubtedly requires a large and expensive labor cost. The Chinese Book Classification Method (CLC), referred to as the Zhongtu Method, is a widely adopted classification standard in China, involving a variety of publications including papers. The zhongtu method is five major categories, twenty-two categories, 51,881 categories (including general categories), there is a hierarchical relationship between each class, and the use of letters and numbers to encode the categories, such as TP181 means "automatic reasoning, machine learning", of which T represents "industrial technology" in the twenty-two categories, TP means "automation technology, computer technology" under the category of "industrial technology", and so on. With the gradual frequency of interdisciplinary intersections, the CLC classification of papers is often composed of multiple components, such as the combination of artificial intelligence and medicine, biology, transportation and other fields has become common.

Research theme is one of the most important applications in the context of big data technology, in recent years, with the rapid development of academic research, all kinds of academic literature showed explosive growth, in the face of massive literature, if you can effectively tap the development trend of research topics, you can provide valuable reference for the research direction of scientific and technological workers. However, predicting trends in research topics has special problems and challenges. Today's various types of academic achievements are growing rapidly, such as deep learning becoming a hot field of artificial intelligence research, which has grown significantly in recent years. However, there are also a large number of traditional research topics that are replaced by new research topics due to their applicability not meeting social changes and poor effects, such as some traditional methods in the field of artificial intelligence, such as genetic algorithms, and the research popularity is decreasing year by year. And the various research topics are not isolated, not only similar research topics have mutual influence, with the gradual deepening of interdisciplinary research, the influence of many different disciplines of research themes is also more close, for example, with the rise of graph neural networks\cite{shi2019deepcollaborativefiltering,zhao2017hinfinityconsensus}, transportation networks began to use graph neural networks to capture spatial features, and the study of smart transportation has gradually become a hot topic. Technologies such as image recognition are widely used in transportation, medical and other multidisciplinary disciplines. Therefore, the traditional sequence model prediction model is simply used, and the dependencies of the research topic cannot be effectively modeled, and the spatial dependencies of the research topic need to be combined.

\section{Semantic Feature Representation and Learning of Scientific Papers}
 Compared with the traditional Internet data, scientific and technological papers show more obvious characteristics, first, scientific and technological data contains both short text and long text, such as keywords and abstract information contained in the paper. Secondly, the relationship between scientific and technological data entities is closely related, such as the citation relationship between papers, the relationship between keywords and papers, the cooperative relationship between scholars, etc., and the effective use of these shallow correlations plays an important role in mining more in-depth paper semantic characteristics\cite{su2020knowledgegraphresources,li2014lpvcontrol}.

TF-IDF ( term frequency -- inverse document frequency) It is a traditional way of extracting text features. It uses word frequency and inverse document frequency, and most represents the document as a multidimensional vector representation of keyword weights, which is a typical vector space model. Mikolov et al. introduce word vector representation model Word2Vec\cite{kim2020word2veclsa}. The entire NLP field soon entered the world of epimediding. The traditional coding method is mainly onehot encoding, the vector obtained by this coding method is often sparse, and the word vector trained by Word2Vec is low-dimensional and dense, which effectively uses the context information of the word, making the semantic information of the vector more abundant. Li\cite{li2020weightedword2vec} et al. used the Word2vec algorithm to handle the semantic divide and implemented a word-frequency-document frequency reciprocal (TF-IDF) weighted mapping of HTTP traffic to construct low-dimensional paragraph vector representations to reduce the complexity of training. However, word2vec, because after retraining, the semantic vector of each word will not change, and it is impossible to obtain different vectors in combination with context semantics.

In 2018, Peters et al. proposed the ELMo model to solve the problem of Word2vec's lack of contextual adaptation\cite{peters2018elmo}. Different from the feature that the semantic representation vector obtained by the static word embedding representation model remains unchanged, ELMo first needs to be pre-trained on a large-scale corpus. Pre-training, so as to achieve the purpose of domain adaptation, so that a word can obtain a special vector according to the current context. The GPT\cite{radford2018gpt} proposed in the same year also uses a corpus to obtain the pre-training model, and then fine-tunes it through a small-scale corpus. Compared with ELMo, the main difference between the two is that the network structure used for feature extraction is different. It is Transformer\cite{vaswani2017attentionarxiv,li2017consensuskalman}, while ELMo uses a more traditional LSTM. Transformer is an end-to-end sequence model proposed by Google. On the basis of this model, many improved methods are widely used in natural language processing, and even in image and other fields. Compared with the traditional sequence model, the Transformer completely adopts the attention mechanism to form the network, and the entire network is composed of the structure of the encoder and the decoder. On this basis, Google further proposed the BERT model\cite{devlin2018bert}. The BERT model uses the mask mechanism to block some words in the corpus for prediction tasks, thereby pre-training the model, and using two-way encoding to effectively extract the context of the text Semantics, breaking records on multiple NLP tasks. Li et al.\cite{li2020contextawarebert} use a gating mechanism with context-aware aspect embeddings to enhance and control BERT representations for perceptual analysis, and TD-BERT\cite{gao2019tdbert} presents three object-dependent variants of the BERT base model, which are described in The output is located on the target word, and optional sentences with the target are built in. Reference\cite{yin2019entitylinking} is based on BERT and designs a negative sampling mining strategy to adjust BERT accordingly. Based on the learned features, I obtain valid entities by computing the similarity between relevant text cues and candidate entities in the knowledge base. BERT4TC\cite{yu2019bertclassification} converts classification tasks into binary pairs by constructing auxiliary sentences, solving the problem of limited training data sensitivity, and the training strategies adopted by BERT for tasks in different domains. Reference\cite{zhang2019jointbert} uses the BERT bidirectional encoder representation to encode the input sequence into a contextual representation. For the decoder, in the first stage, the intent classification decoder is used to detect the user's intent. In the second stage, the intent context information is exploited into the slot filling decoder to predict the semantic concept labels for each word. Recent work also shows that retrieval-oriented pre-training can further improve representation learning for downstream search and matching. For example, RetroMAE pre-trains retrieval-oriented language models through a masked auto-encoder objective, which is useful for representing scientific papers in retrieval and recommendation scenarios\cite{xiao2022retromae}.

Beyond word- and sentence-level semantic representation, scientific paper representation also needs to model structured relations among papers, authors, venues, keywords, and citations. Heterogeneous graph attention networks have been used for semi-supervised short text classification, showing that heterogeneous relations among words, documents, and semantic contexts can alleviate the sparsity of short text features\cite{hu2019hgat}. Li et al. proposed a semantic-similarity attention and hypergraph convolution framework for scientific publication representation learning, which explicitly captures semantic similarity and high-order associations among publications. This line of work indicates that hypergraph-based representation learning can complement Transformer-style text encoders when mining the semantics of scientific papers\cite{li2026ssahgc}.

\section{Domain Information Mining of Scientific Papers}
 Mining the semantics of the subject area from the paper, that is, to associate the scientific and technological entities with the subject area first. With the continuous development of science and technology\cite{li2022smcr}, the cross-integration between disciplines is becoming more and more frequent. For example, artificial intelligence has penetrated into all aspects of life. The subject area semantics of scientific and technological entities is no longer a single subject area, and often one entity is closely related to multiple subject areas. This can thus be translated into a multi-label classification problem for papers, associating papers with multiple subject area labels.

Scientific and technological information may also appear in multiple media forms. Li et al. proposed a semantics-adversarial and media-adversarial cross-media retrieval method for scientific and technological information, which learns a common subspace while preserving semantic consistency and media invariance\cite{li2022smcr}. More recently, federated supervised cross-modal retrieval has further shown how distributed multimodal resources can be exploited without directly centralizing raw data, providing a useful privacy-aware direction for mining heterogeneous scientific and technological information\cite{li2024fedcmr}.

There are three main approaches to traditional multi-classification problems. The first is a problem transformation-based approach, where for each label, a separate binary classifier is trained separately for prediction, which increases computational complexity while dealing with a large number of labels. References\cite{yen2016pdsparse,yen2017ppdsparse} use L1 regularization to achieve sparse solutions, thus reducing the computational complexity. Babber et al.\cite{babbar2016dismec} learn a linear classifier for each label and use two-layer parallelization to control the model size. Another obvious limitation of this approach is the ignorance of label dependencies during training, which may weaken the generality of the model. To address this limitation, a Bayesian network is employed to encode label dependencies by computing the joint probability of labels and feature sets. The second is based on decision tree models, which are inspired by the idea of decision trees and build decision trees based on labels or data instances by recursively splitting internal nodes. Predictions are made before new data points are passed down the tree until they reach a leaf node. FastXML achieves remarkable accuracy by directly optimizing the ranking loss function. PfastreXML\cite{jain2016extrememultilabel} is an extension of FastXML that prioritizes the prediction of tails and handles missing tags by proposing a propensity score loss. Parabel\cite{prabhu2018parabel} generates an ensemble of balanced trees on labels rather than data points, which can be seen as using softmax to improve performance. Embedding-based methods employ a compression function to project label embeddings into a lower-dimensional linear subspace. SLEEC\cite{bhatia2015sleec} was proposed to address the limitation that low-rank label matrix assumptions are often violated in real-world applications. AnnexML\cite{tagami2017annexml} proposes a graph embedding method to cope with some limitations of SLEEC.

With the gradual improvement of the performance of deep networks\cite{xu2013imagefusion,lin2009averageconsensus} , attempts to use deep methods for multi-label classification have increased in recent years. Liu et al.\cite{liu2017deepxml} adapted the MTC task by adding dynamic max pooling to capture more fine-grained information and using bottleneck hidden layers to reduce parameter size. Nam\cite{nam2017subsetaccuracy} converts the predicted values of a set of related labels into a predicted sequence of binary values and uses a recurrent neural network (RNN) for prediction. Yeh\cite{yeh2017latentmultilabel} proposed to adopt a norm-dependent auto-encoder to jointly model text features and label structures. Zhang et al.\cite{zhang2017deepxml} constructed a label graph in an attempt to explore the label space. Wang et al.\cite{wang2018jointembedding} learned joint word label embeddings and combined the context vectors into one final document vector using the compatibility score between each label and context words as attention coefficients. Similarly, You et al.\cite{you2019attentionxml}. used multiple label attention vectors to allow the network to incorporate multiple semantics for each label. It produces multiple label attention vectors, and each label vector maps to a single output. Tang et al.\cite{tang2015pte} established a textual heterogeneous network to encode multi-level semantic information. Graph Convolutional Networks have gradually gained popularity in multi-class classification. Rousseau\cite{rousseau2015textgraph} and Yao\cite{yao2019textgcn} use GCNs to jointly learn word and document embeddings for graphical text representations.

Classification task is a traditional task of artificial intelligence. However, in reality, a large number of transactions exist with multiple labels. The purpose of multi-label classification is to give a sample and obtain the label set of the sample. The association between labels can be exploited in multi-label classification tasks. Zhang et al.\cite{zhang2020featureselection} proposed a global optimization method, which aims to consider feature correlation, label correlation and feature redundancy for feature evaluation. Feng et al.\cite{feng2019collaboration} learned label correlations through sparse reconstruction in the label space and incorporated the learned label correlations into model training. Many labels lack sufficient samples, and Lv et al.\cite{lv2019labelenhancement} use structural information in the feature space and local correlations in the label space to enhance the labels. Xing et al.\cite{xing2018cotraining} exploited information about the coexistence of pairwise labels to propagate the labels of selected samples among co-trained classifiers. Xun et al.\cite{xun2020correlationnetworks} developed a correlation network architecture to learn label correlations, use correlation knowledge to enhance the original label predictions and enhance the output label prediction results. Shi et al.\cite{shi2020crowdsmultilabel} proposed a deep generative model to generate labels for semi-supervised learning by combining latent variables to describe labeled and unlabeled data. In practical field information mining, the interpretability of classification and decision models is also important, because interpretable machine learning can provide explainable evidence for intelligent decision-making over complex scientific and technological entities\cite{li2019interpretable}.

In recent years, graph learning has developed rapidly, and Velikovi and Petar et al. used an attention mechanism to calculate the weights of different nodes in the neighborhood without relying on the global structure of the graph. Yao et al.\cite{yao2019textgcn} build a single text graph for the corpus based on word co-occurrences and document word relationships, and then learn a text graph convolutional network. Based on GraphSage, Tang et al.\cite{tang2020patentgcn} obtained second-order features using BiLSTM as an aggregation function to capture dependencies. Wang et al.\cite{wang2020labelgraph} model the label graph with co-occurrence information, and then apply multi-layer graph convolution on the final overlay graph for label embedding. Community detection and incomplete-graph representation are also relevant to scientific-paper networks: modularity-based deep community detection can help reveal scholar or citation communities, while T2-GNN studies feature- and structure-incomplete graphs through teacher-student distillation, which is suitable for noisy or sparse bibliographic networks\cite{yang2016modularity,huo2023t2gnn}.

Federated graph and information-network representation learning provides another direction when scientific data are distributed across institutions. Federated graph neural networks for cross-graph node classification show how node representation can be learned across multiple graphs without directly sharing all raw graph data\cite{guan2021fedgnn}. FedSIN further models non-Euclidean information networks in a federated setting by combining graph attention with self-adaptive client aggregation, which is relevant for paper, keyword, author, and citation networks that cannot be easily centralized\cite{li2026fedsin}.

Unlike traditional multi-label classification tasks, in hierarchical multi-label tasks the labels are organized into a hierarchy. Considering that conceptual relationships between words can also form hierarchies, Chen et al.\cite{chen2020hyperbolic} mapped from word hierarchies to label hierarchies. Wehrmann et al.\cite{wehrmann2018hierarchical} utilize multiple linear layers (corresponding to the number of category layers) and have local outputs in each layer. It optimizes the loss of the local layers and the overall loss of the final output. Yan et al.\cite{yan2018activelearning} combined the potential contributions of parent labels to child labels to evaluate the confidence of each label.

\section{Analysis of the Evolution Law of Research Topics in Scientific and Technological Papers}
 As a more complex relational data structure that can describe entities, graphs can effectively represent the spatial dependencies of research topics, and the representation learning of graph states\cite{zhang2020featureselection,li2017stateestimation,li2017recursiveestimation} has gradually attracted attention. In recent years, with the rise of deep learning, new breakthroughs have been made in graph learning (such as GCN\cite{kipf2016gcn} , GAT\cite{velickovic2017gat} , Walk Pooling\cite{pan2021walkpooling}), and people have developed great interest in graph representation learning. The above figure shows that the learning work is only concentrated on the static graph, that is, the node characteristics and graph structure of the graph will not change. However, there are lots of dynamically changing graphs. For example, the transmission path of the virus will change over time, and the number of infections at various locations will also change. Reflected in the graph, the structural features and node features of the graph will change dynamically according to time. Effective prediction of local epidemics plays a crucial role in virus protection\cite{panagopoulos2021pandemicforecasting,hu2018anomalydetection}. In social networks\cite{li2021communitydiscovery,kou2016socialnetworksearch}.people's social relationships are also dynamically changing, and user behavior is constantly changing, so the vector representation of users should be updated accordingly. Likewise, the citation network of scientific articles is constantly enriched due to the frequent publication of new works citing existing technologies. Therefore, the popularity and influence of some technologies change over time\cite{zhou2021academicinfluence,li2013gmphd}.

In scholarly data analysis, Li et al. studied multi-view scholar clustering with dynamic interest tracking, showing that scholars should be modeled through multiple academic views and temporally evolving interests rather than as static entities\cite{li2023mvsc}. This perspective is consistent with the need to model dynamic academic entities and evolving topic relations in research-topic evolution analysis.

With the rapid development of deep learning, recurrent and recursive network structures have begun to replace traditional linear models such as (ARIMA), which are widely used in sequence problems. However, models such as RNN are difficult to extract long-distance features effectively. Qin et al.\cite{qin2017darnn} propose a two-stage attention-based recurrent neural network, and on the basis of the attention mechanism, the LSTM structure was used to extract the long and short distance features of the time series. LSTNet\cite{lai2018lstnet} employs Convolutional Neural Networks (CNN) and Recurrent Neural Networks (RNN) to extract short-term local dependency patterns among variables and discover long-term patterns of time series. The Transformer\cite{vaswani2017attentionnips} network structure proposed by Google uses a pure attention mechanism and has achieved significant improvements in most sequence problems, but there is a problem of not being able to highlight local information. Since in the Transformer, the token only reflects the distance relationship through the position information, and theoretically all its nodes are equal, so the local information cannot be reflected. R-Transformer\cite{wang2019rtransformer} first uses RNN to model local information, and then inputs it into Transformer, which avoids the problem of equalization of local information and global information. However, these network structures are difficult to parallelize, and the amount of computation is large. Related sequence modeling work in recommendation, including filter-enhanced MLP and self-supervised graph co-training for session-based recommendation, also suggests that temporal user-item interactions can be modeled by non-recurrent filters or self-supervised graph signals; these methods can serve as auxiliary references for modeling dynamic topic or scholar-interaction sequences\cite{zhou2022fmlp,xia2021graphcotrain}. With the widespread application of convolutional neural networks\cite{xue2019lowrankensemble,fang2020identitycyclegan} in the field of two-dimensional images. In order to solve the problem that it is difficult for convolutional networks to capture long-distance features, Bai et al.\cite{bai2018tcn} uses dilated convolutional and fully convolutional networks to significantly improve the receptive field of the network, and proves that compared with LSTM/GRU and other models, the efficiency and accuracy are improved. He et al.\cite{he2022grulstm} combines the advantages of high prediction accuracy of LSTM and short prediction time of GRU to efficiently predict cloud computing resource load. Variance-constrained estimation in nonlinearly coupled complex networks also provides a useful reference for modeling uncertainty in dynamic graph states\cite{li2017variancecybernetics}.

To address the problem of representation of spatial features in prediction tasks, T-GCN\cite{zhao2019tgcn} combines Graph Convolutional Networks (GCNs) and Gated Recurrent Units (GRUs). Among them, GCN is used to learn complex topological structures and capture spatial correlations; Gated Recurrent Unit (GRU) is used to learn dynamic changes of traffic data and capture temporal correlations. Then, the T-GCN model is used for traffic prediction based on the urban road network. A3TGCN\cite{guo2019astgcn} utilizes spatiotemporal attention mechanism to learn dynamic spatiotemporal correlations of traffic data and combines spatiotemporal convolution for traffic prediction. LRGCN\cite{li2019lrgcn} treats temporal dependencies between temporally adjacent graph snapshots as a special relation to memory, and uses relational GCNs to jointly handle temporal and temporal relations. AGCRN\cite{bai2020agcrn} employs a Node Adaptive Parameter Learning (NAPL) module to capture node-specific patterns, and a Data Adaptive Graph Generation (DAGG) module to automatically infer the interdependencies between different flows. The GCN in GC-LSTM\cite{chen2021gclstm} is capable of node structure learning for each time-slipped network snapshot, while the LSTM is responsible for the temporal feature learning of network snapshots. Yu et al.\cite{yu2021ridedemand} model the correlation between future demand and space-time through DCN and LSTM, and modeled the demand to predict the regional ride demand. Li et al.\cite{li2021dynamicgraphgru} use an autoencoder as a framework, in which the encoder first used DNN to aggregate the neighborhood information to obtain low-dimensional feature vectors, then used the GRU network to extract node temporal information, and finally used the decoder to reconstruct the adjacency matrix and put it into Comparing with the real graph to construct the loss. Sun et al.\cite{sun2022weightedpointcloud} design an appropriate weighting function for the edge features composed of k-nearest neighbor graphs\cite{sun2009knn} to weaken the interference of far points, relatively strengthen the features of near points, and use a symmetric function that combines maximum pooling and average pooling to compensate global information loss.

\section{Conclusion}
 Aiming at the research on the field information mining\cite{yang2015ontologyretrieval} and theme evolution law of scientific and technological papers, this paper expounds the research status at home and abroad from three aspects: the semantic feature extraction of scientific and technological papers, the field information mining of scientific and technological papers, and the research on the theme evolution law of scientific and technological papers. With the development of natural language processing and other fields, the research on scientific papers has borrowed a lot of theories and methods of natural language processing. We think that there are still challenges in the following aspects. In the research on the evolution law of research topics, it has been through dynamic The method of graph learning realizes the function of predicting the number of future results of each research topic, but the correlation prediction between research topics still needs to be studied. The size of the paper, the lack of effective means to deal with large-scale paper datasets. Most of the current methods conduct experiments on a small number of samples, facing the huge scale of paper data, lack of effective means for processing large-scale paper datasets.

\begin{acks}
This work was supported by National Key R\&D Program of China (2018YFB1402600), and by the National Natural Science Foundation of China (61802028, 61772083, 61877006, 62002027).
\end{acks}

\bibliographystyle{unsrt}
\bibliography{references}
\end{document}